# Optimal tree for Genetic Algorithms in the Traveling Salesman Problem (TSP).

Liew Sing
liews_ryan@yahoo.com.sg

April 1, 2012

**Abstract**
In this paper, the author proposes optimal tree as a "gauge" for the generation of the initial population at random in the Genetic Algorithms (GA) to benchmark against the good and the bad parent tours. Thus, without having the so-called bad parent tours in the initiate population, it will speed up the GA. The characteristics of the gauge (algorithm, complexity time, trade-off, etc.) will be discussed in this paper as well.

**1.Introduction**
Despite the fact that the Traveling Salesman Problem (TSP) is very intuitive and easy to state, it is one of the most widely studied NP-hard combinatorial optimization problem[14]. The following are the statements of the problem. A salesman is required to visit each of *n* given cities once and only once, starting from any city and returning to the original city of departure. How should he travel in order to minimize the total travel distance?[8] The difficulty becomes obvious when one considers the number of possible tours by the method of brute force searching even for a relatively small number of cities *n*. For instance, for a problem with 20 cities (*n*=20) by brute force searching, it would be (20-1)!/2 tours, which is more than $10^{18}$ tours! The TSP is a class of difficult problems whose time complexity is widely believed exponential. Any attempt to construct an algorithm for finding optimal solutions for the TSP in polynomial time (in contrast with exponential time) is also widely believed not possible. Up to date, there are two classes of algorithms in solving the TSP: *Exact* algorithms and *Approximate* (or *heuristic*) algorithms[8]. The main characteristics of Exact algorithms are guaranteed to find the optimal solution in a bounded number of steps but unfortunately also complex with codes and very demanding of computer power[8]. Examples of the most effective Exact algorithms are Cutting-Plane and Facet-Finding algorithms[8]. On the other hand, in contrast, the main characteristics of Approximate algorithms are no guarantee that optimal solutions will be found but nevertheless able to provide relatively good solutions (differs only by a few percent from the optimal solution) and these algorithms are usually have shorter running times and very simple[8]. There are three classes of Approximate algorithms[8]: 1) Construction algorithms e.g. Nearest Neighbor algorithms, which gradually construct a tour by adding a new city at each step. 2) Improvement algorithms e.g. Genetic Algorithms (GA)[6]. And 3) Composite algorithms, which combine Construction and Improvement algorithms[8][16].

As a matter of fact, GA is one of the fastest and widely adopted method in searching for the optimal tour in the TSP[6]. However, at the beginning of the algorithm, which generates the initial population at random, the author thinks that the algorithm should include a "gauge" to benchmark against the good and bad tours which were created at random. With the gauge, the initial population will has only the good parent tours and thus eventually speed up the algorithm searching for the optimal tour tremendously. The sequence of the paper is as follows. Firstly, the author will briefly discuss about the GA. Secondly, we will look into the optimal tree as well as its algorithms. Lastly, the paper ends with discussions and conclusions.

## 2. Genetic Algorithms (GA)

Genetic Algorithms (GA) is one of the branches of Evolutionary Algorithms (EA), which are probabilistic search algorithms simulate natural evolution. EA was inspired by Darwin's "survival of the fittest" and the evolution theory of natural selection that based on the claim that, in nature, there exist many processes which always seek a stable state. In other words, these processes can be described as natural optimization processes[13]. In addition, other than GA, there are a few more branches of EA. Namely, evolution strategies[2], classifier systems[10], genetic programming[11], evolutionary programming[5], etc. EA were proposed more than 40 years ago[6][13] and GA were introduced by Holland in 1975[10]. However, it was in the 90s that applying GA to the TSP became an actual research topic.

Technically, GA operates with a large number of tours in the sense that it produces the initial population of tours at random and consecutively several other populations such a way that the best tour in the current populations is not worse than the best tour in the initial population[6]. Figure 1 illustrates the algorithms of GA in the TSP and it works as follows[13].

```
BEGIN
    Make initial population at random.
        WHILE NOT stop DO
            BEGIN
                Select parents from the population.
                Produce children from the selected parents.
                Genetic operator acts on the children.
                Extend the population adding the children to it.
                Reduce the extend population.
            END
    Output the best individual found.
END
```

Figure 1: Genetic Algorithms (GA) for the TSP.

Firstly, the initial population is chosen and their quality is determined and measured with an evaluation function. Secondly, parents are selected from the population to produce children. The children will be genetically altered by the *genetic operator* and added to the population. Next, some individuals will be removed from the population based to the evaluation function in order to maintain the population size. One iteration of the algorithms of the *Begin* to *End* of the inner loop is called *generation*. Lastly, after rounds of generations, the best tour will be chosen as the optimal tour. In step two, there are two genetic operators that will improve the child tour available so far: *mutation* and *crossover*. A mutation means making small changes to the child tour; crossover will identify a set of edges *E(A)* from the parent tour *A* and another set of edges *E(B)* from the parent tour *B* such a way that these sets of edges form another set of *smaller tours*. These smaller tours will be implemented on the parent tours and thus created some better child tours[13][14].

**3. Optimal tree**

An optimal tree is a minimum spanning tree. A tree means a connected graph with no cycles in it and if the tree reaches out to all the vertices, it will be a spanning tree. Suppose we have a weighted (or so-called *cost*), undirected (edges have no orientation, e.g. edge of *ab* = edge of *ba*) and connected graph $G=(V,E)$, where $V$ is a finite set of vertices (or points) $v$ and $E$ is a finite set of edges (or paths) $e$. In addition, there is a weight function of edge $w(e)$. So, an optimal tree $T$ will be $w(T) = min \sum_{e \in T} w(e)$ as shown in Figure 2[3].

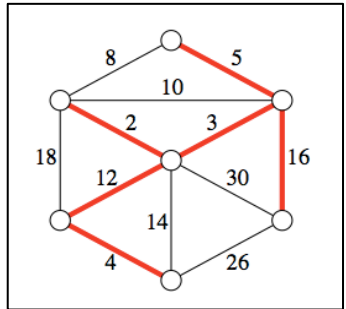

Figure 2: A weighted graph with different weights and its optimal tree (in red).

In general, there are three algorithms searching for optimal tree, namely Boruvka's algorithm (BA)[1], Prim's algorithm[15] and Kruskal's algorithm[12]. All these algorithms, however, are based on greedy algorithms and the complexity times are all $O(E\log V)$, which are "polynomial"[3]. Since all these algorithms sharing the same complexity time, the author would like to have only BA included in this paper. In addition, BA is arguably the simplest as well. Reader who are interested in Prim's algorithm and Kruskal's algorithm may which to seek their original paper [15] and [12], respectively or textbook [17].

Intuitively, all the methods for computing optimal tree are so-called generic (not *genetic*) algorithm that will add (or merge) trees together by adding certain edges between them. The generic optimal tree algorithm maintains an acyclic subgraph of the input graph $G$. We call the acyclic subgraph an *intermediate spanning forest F* and it is a subgraph of the optimal tree of $G$, that is, every component of $F$ is an optimal tree of its vertices. Thus, $F$ induces two types of edges: *useless* edge and *safe* edge. A useless edge is not an edge of $F$ but both its vertices are in the same component $F$. For each component, a safe edge is the minimum-weight edge with exactly one vertex in that component. On the other hand, the edges that are neither safe nor useless are called *undecided.* So, it is easy to see that an optimal tree will contain every safe edge and has no useless edges. Technically, generic optimal tree algorithm will repeatedly add one or more safe edges to the evolving $F$ and make some undecided edges become safe. Likewise, some edges become useless. By the same token, BA is basically an algorithm that keeps searching for the safe edges and adds them. Figure 3 illustrates BA with the weighted graph given previously in Figure 2[3]. Figure 4 illustrates the algorithm of BA[9].

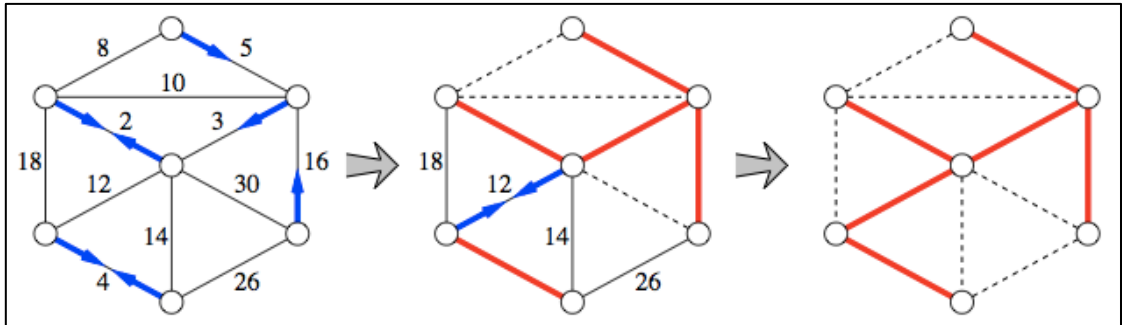

Figure 3: Dashed edges are useless edges and arrows point along each component's safe edge

```
Given G = (V,E)
Declare T = graph consisting of V with no edges
while T has < n-1 edges do
    for each connected component C of T do
            e is equal to the minimum weight edge (v,u) such that v in C and u not in C
            T define as T union {e}
```

Figure 4: The algorithms of BA.

This proposal of suggesting optimal tree as a "gauge" for the generation of the initial population at random in the Genetic Algorithms (GA) to benchmark against the good and the bad parent tours was inspired by the fact that, in general, an optimal tree contains many edges in common with an optimal tour of the TSP[8]. That is, any parent tour generated for the initial population that has many edges in common with the optimal tree will give advantages to the GA both in complexity time and complexity space. The advantage in complexity time will be the increase of speed of the GA because, recall the algorithms of GA in Figure 1, we can get rid of the algorithm of "Select parents from the population" of the inner loop to save some times for a generation and, collectively after a lot of generations, save a lot of times. The advantage in complexity space will be the saving of the memory resource to include those bad parents tours which unlikely to produce better children tour in comparison with those from the good parents tours. However, the trade-off of the gauge will be the requirement of running the algorithms of searching for the optimal tree. Nevertheless, the author finds it should not be a problem because we can search for the optimal tree on another computer (later implement the optimal tree into the GA's computer) and does not affect the overall GA for the TSP. Therefore, we have a new modified algorithms for the GA and it is shown in Figure 5.

```
BEGIN
    Make initial population based on the gauge given
        WHILE NOT stop DO
            BEGIN
                Produce children from any parents.
                Genetic operator acts on the children.
                Extend the population adding the children to it.
                Reduce the extend population.
            END
    Output the best individual found.
END
```

Figure 5: New Genetic Algorithms (GA) for the TSP.

## 4. Discussions and Conclusions

In 1969, Held and Karp published a paper, titled "The Traveling-Salesman Problem and Minimum Spanning Trees"[7], claiming that a) a tour is precisely a 1-tree in which each vertex has degree 2, b) a minimum 1-tree is easy to compute and c) the transformation on "intercity distances" $c_{ij} \to c_{ij} + \pi_i + \pi_j$ leaves the traveling-salesman problem invariant but changes the minimum 1-tree. Technically, 1-tree is a graph with vertices $v_1, v_2, \ldots, v_n$ consisting of a tree on the vertices $v_2, v_3, \ldots, v_n$ together with two edges incident with $v_1$ i.e. a 1-tree has a single cycle, this cycle contains vertex $v_1$ and $v_1$ always has degree two. Later on, in 2000, Helsgaun also claimed that, in general, an optimal tour contains between 70% and 80% of the edges of minimum 1-tree[8] which is obviously important and significant in our study here. Claim a) is obvious and trivial for our study here. For claim b), Held and Karp suggested a simple algorithm[7]: a minimum-weight 1-tree can be found by constructing a minimum spanning tree (optimal tree) on the vertex set $V = \{v_2, v_3, \ldots, v_n\}$ without $v_1$, and then adjoining two edges of lowest weight at $v_1$. So, in our algorithms of BA, it would mean we have only need to change $V = \{v_1, v_2, \ldots, v_n\}$ to $V = \{v_2, v_3, \ldots, v_n\}$ and then we may use Nearest Neighbor to adjoin two edges at $v_1$. As for claim c), it is still unclear whether this transformation would help in our proposal or not. But apparently, this transformation looks promising in improving the percentage of having the edges common in minimum 1-tree and the optimal tour because, by collecting different minimum 1-trees and perhaps putting them together, it would produce an even better gauge and thus save us even more complexity time and complexity space.

On the other hand, the reliability of randomized algorithms has sparked a debate in the communities of mathematics and computer science recently. Some computer scientist argued that randomized algorithms in general are not "random" at all and it actually follows a *deterministic* fashion which depends on earlier results[4]. Since GA depends on making initial population at random, the reliability of GA is in doubt as well. The success of GA, particularly for the TSP, is actually relies on *randomness* or *deterministic* fashion? We are able to obtain the optimal tour in a fast manner is due to the fast and effective genetic operators or due to the deterministic fashion which always create good parent tours (because we have created a good parent tour at the first place) for the initial population? What if the initial population was created based on a bad parent tour at the first place? Because of the uncertainty of the reliability of randomized algorithms, it would require the TSP researchers whose work are based on GA to perform their experiment as many times as possible in order to obtain the *mean* and have a complete understanding of the genetic operators. It is the genetic operators of *edge assembly crossover* (EAX) that make the GA as one of the most efficient algorithms in the TSP[6]. Unfortunately, it is not practical because it may takes weeks or months to run the algorithms to search for an optimal tour which has large number of *n* cities. However, with the implementation of the gauge in the GA, computer scientists not only able to increase the efficiency of the GA, the series of questions can be avoided as well and thus focus on the study of genetic operators without worrying about the initial population because all parent tours are good tours. Or we can actually set the gauge to isolate the good parent tours then we can see the "real" efficiency of our genetic operators after rounds of generations. On the other hand, we can also use the gauge to tell the computer to pick good and bad tours alternately so that we can have "real random" initial population.

In conclusion, the author proposed optimal tree (or minimum spanning tree) as a gauge to isolate the bad parent tour to include in the initial population for the Genetic Algorithms in the TSP. By doing so, it would save a lot of complexity time as well as complexity space. The author also provides new algorithms for the Genetic Algorithms to suit the implementation of the gauge. The author also raised the concern about the reliability of randomized algorithms which play an important role in the GA. Last but not least, the author also suggest some further researches on minimum 1-tree, particularly on the transformation of the TSP's "intercity distances" $c_{ij} \to c_{ij} + \pi_i + \pi_j$ to increase the percentage of having the edges common in minimum 1-tree and the optimal tour.